\documentclass[fleqn,usenatbib]{mnras}

\usepackage{newtxtext}
\usepackage[varvw]{newtxmath}

\usepackage[T1]{fontenc}

\DeclareRobustCommand{\VAN}[3]{#2}
\let\VANthebibliography\thebibliography
\def\thebibliography{\DeclareRobustCommand{\VAN}[3]{##3}\VANthebibliography}

\usepackage{graphicx}	% Including figure files
\usepackage{amsmath}	% Advanced maths commands
\graphicspath{{Images/}}

\title[Tidal evolution of the Fornax dSph]{The tidal evolution of the Fornax dwarf spheroidal and its globular clusters}

\author[A. Borukhovetskaya et al.]{
Alexandra Borukhovetskaya$^{1}$\thanks{E-mail: asya@uvic.ca},
Rapha\"{e}l Errani$^{1,2}$,  
Julio F. Navarro$^{1}$,
Azadeh Fattahi$^{3}$,
Isabel Santos-Santos$^{1}$
\\
$^{1}$Department of Physics and Astronomy, University of Victoria, Victoria, BC V8P 5C2, Canada\\
$^{2}$ Observatoire Astronomique, Universit\'e de Strasbourg, CNRS, 11 rue de l'Universit\'e, 67000 Strasbourg, France\\
$^{3}$ Institute for Computational Cosmology, Department of Physics, University of Durham, South Road, Durham DH1 3LE, UK
}

\date{Accepted XXX. Received YYY; in original form ZZZ}

\pubyear{2020}

\begin{document}
\label{firstpage}
\pagerange{\pageref{firstpage}--\pageref{lastpage}}
\maketitle

% Abstract of the paper
\begin{abstract}
The dark matter content of the Fornax dwarf spheroidal galaxy inferred from its kinematics is substantially lower than expected  from LCDM cosmological simulations.
We use \textit{N}-body simulations to examine whether this may be the result of Galactic tides. We find that, despite improved proper motions from the Gaia mission, the pericentric distance of Fornax remains poorly constrained, mainly because its largest velocity component is roughly anti-parallel to the solar motion. Translating Fornax's proper motion into a Galactocentric velocity is thus sensitively dependent on Fornax's assumed distance: the observed distance uncertainty, $\pm 8\%$, implies pericentric distances that vary between $r_{\rm peri}\sim 50$ and $r_{\rm peri}\sim 150$\,kpc. Our simulations show that for $r_{\rm peri}$ in the lower range of that estimate, a LCDM subhalo with maximum circular velocity $V_{\rm max}=40$\,km\,s$^{-1}$ (or virial mass $M_{200}\approx 10^{10}\, M_\odot$, as expected from LCDM) would be tidally stripped to $V_{\rm max} \sim 23$\,km\,s$^{-1}$ over $10$\,Gyr.  This would reduce the dark mass within the Fornax stellar half-mass radius to about half its initial value, bringing it into agreement with observations. Tidal stripping affects mainly Fornax's dark matter halo; its stellar component is affected little, losing less than $5\%$ of its initial mass in the process. We also explore the effect of Galactic tides on the dynamical friction decay times of Fornax's population of globular clusters (GC) and find little evidence for substantial changes, compared with models run in isolation. A population of GCs with initial orbital radii between $1$ and $2$\,kpc is consistent with the present-day spatial distribution of Fornax GCs, despite assuming a cuspy halo. Neither the dark matter content nor the spatial distribution of GCs of Fornax seem inconsistent with a simple model where Fornax inhabits a tidally-stripped cuspy cold dark matter halo.\end{abstract}


%%%%%%%%%%%%%%%%%%%%%%%%%%%%%%%%%%%%%%%%%%%%%%%%%%

%%%%%%%%%%%%%%%%% BODY OF PAPER %%%%%%%%%%%%%%%%%%

\section{Introduction}
\label{SecIntro}

\begin{table*}
	\centering
	\caption{Parameters of the analytical, static Milky Way potential used in this study. The model is a spherical re-parametrisation of the \citet{McMillan2011} model, as discussed in \citet{Errani2020}.}
	\label{tab:MW}
	\begin{tabular}{lllll}
		\hline
		Component & Functional form & & & \\
		\hline
		Disk (thin) & \citet{Miyamoto1975} & $M=5.9\times10^{10}\,\rm M _\odot$ & $a_{\rm d} = 3.9$\,kpc & $b_{\rm d} = 0.31$\,kpc \\
		Disk (thick) & \citet{Miyamoto1975} & $M=2.0\times10^{10}\,\rm M _\odot$ & $a_{\rm d} = 4.4$\,kpc & $b_{\rm d} = 0.92$\,kpc \\
		Bulge & \citet{Hernquist1990} & $M=2.1\times10^{10}\,\rm M _\odot$ & $a=1.3$\,kpc & \\
		DM Halo & \citet{Navarro1997} & $M_{200}=1.15\times10^{12}\,\rm M _\odot$ & $r_{\rm s} = 20.2$\,kpc & $c=r_{200}/r_{\rm s}=9.5$\\
		\hline
	\end{tabular}
\end{table*}

The standard model of cosmology, Lambda Cold Dark Matter (LCDM), predicts structures to form hierarchically. In this paradigm, large self-bound structures (haloes) form through the accretion and mergers of smaller ones (subhaloes), which in turn form from small scale density perturbations in the early Universe \citep{White1978}. Massive haloes allow stars to form in their centres from cooling of accreted gas: the more massive the host halo, the more stars it is able to form. In low-mass haloes ($\lesssim 10^9 \rm M_\odot$), however, cosmic reionization and energetic feedback from stellar evolution reduce substantially a halo's ability to form stars \citep{Bullock2000,Benson2002,Somerville2002}.

This implies that a steep non-linear relation between the dark matter masses of the smallest haloes and their stellar masses is expected in LCDM, a prediction that has been supported by both semianalytic techniques such as "abundance-matching" \citep{Guo2010,Moster2013,Behroozi2013}, and by direct cosmological hydrodynamical simulations \citep[e.g., the Illustris and EAGLE simulations,][respectively]{Vogelsberger2014, Shaye2015}.

More specifically, on the scale of dwarf galaxies, simulations predict that galaxies with stellar masses $M_{\rm str} \sim 2.4\times 10^7 \, M_\odot$, like the Fornax dwarf spheroidal (dSph) (see Table \ref{tab:Fornax_structure}), should form in a halo with a virial\footnote{We define the virial mass as the mass enclosed within the radius of overdensity 200 times the critical density required for closure, $\rho_\mathrm{crit}={3H^2/8\pi G}$, where $H(z)$ is Hubble's constant and $H_0 =H(0)= 67.74$\,km\,Mpc$^{-1}$\,s$^{-1}$ \citep{Planck2015}} mass of at least $M_{200}\approx 10^{10}$\,M$_\odot$ \citep{Wang2015,Fattahi2018,Garrison-Kimmel2019,Munshi2021}.

In LCDM, where the dark matter halo mass distribution is adequately approximated by Navarro-Frenk-White profiles \citep[hereafter NFW]{Navarro1996b,Navarro1997}, this virial mass corresponds to a halo with maximum circular velocity, $V_{\rm max}=39.6$\,km\,s$^{-1}$, for a "concentration" parameter $c=12.5$ \citep{Ludlow2016}. The mass profile of such a halo is fully specified, enabling predictions of the dark mass enclosed within the galaxy stellar half-mass radius\footnote{We shall use uppercase $R_{1/2}$ to denote projected half-mass radii and lowercase $r_{1/2}$ to denote deprojected, 3D, radii. We shall usually assume $R_{1/2}=(3/4)\, r_{1/2}$, as appropriate for spherical systems, unless otherwise noted.}, where observational constraints are tightest  \citep{Walker2009, Wolf2010}. In the case of the Fornax dSph, $r_{1/2}\approx 1$\,kpc, and observations suggest a total enclosed mass of $M_{\rm tot}(<r_{1/2}) \approx 9 \times 10^7\, M_\odot$ (or, equivalently, a circular velocity of $V_{c}(r_{1/2})\approx 20$\,km\,s$^{-1}$), much larger than the mass in the stellar component \citep{Fattahi2016b,Read2019}.

For an NFW halo of average concentration, the latter constraint implies $V_{\rm max} \sim 25$\,km\,s$^{-1}$, substantially below that expected from the cosmological simulations referenced above. The disagreement is amplified by the steep non-linear dependence of stellar mass on halo mass in this regime. Indeed, in LCDM haloes with $V_{\rm max}\sim 25$\,km\,s$^{-1}$ are expected to harbour dwarfs with $M_{\rm str}\sim 10^6\, M_\odot$, more than an order of magnitude less luminous than Fornax. If such low-mass haloes were to harbour galaxies as massive as Fornax, one would expect nearly an order of magnitude more Fornax-like dwarfs than observed in the Local Group.

The unexpectedly low dark matter content of Fornax is often cited as suggesting the presence of a constant-density ``core'' in the inner halo density profile \citep[see, e.g.,][]{Walker2011,Amorisco2013}. Such cores may result from the supernova-driven cycling of baryons in and out of the inner regions of a halo during galaxy formation  \citep[see; e.g.,][]{Navarro1996b, ReadGilmore2005, PontzenGovernato2012, DiCintio2014}, but their occurrence is not universally accepted. %%For example, there is no evidence of baryon-induced cores in otherwise successful galaxy formation simulations with similar resolution to aforementioned works, like those of the Illustris and EAGLE collaborations \citep[see; e.g.,][]{Bose2019}.

In the absence of cores, one would need to argue that the dark matter content of Fornax has been eroded somewhat by Galactic tides. Fornax's stellar component shows no obvious sign of tidal disturbance, however, but tidal stripping is expected to affect mainly the more extended dark matter component \citep[see; e.g.,][]{Penarrubia08}. Tidal stripping may thus lead to a reduction of the dark matter content of a dwarf without affecting much the stellar component. This has been argued by \citet{Genina2020}, for example, who selected dwarf galaxies with orbital properties similar to Fornax in the APOSTLE simulations, and showed that many such satellites lose more than half of their mass from within their innermost $\sim 1$\,kpc.

Although plausible, the main obstacle to this tidal-stripping interpretation is Fornax's large Galactocentric distance \citep[$\sim 149$\,kpc][]{Pietrzynski2009}, together with early estimates of its orbital eccentricity, which suggested a low-eccentricity orbit \citep[see; e.g.,][and references therein]{Piatek2002,Dinescu2004,Battaglia2015}. At such distance, it would be unlikely for Fornax to have been affected much by Galactic tides if it was on a nearly circular orbit.

Besides its unexpectedly low dark matter content, Fornax has long been argued to pose an additional problem for LCDM. This relates to its globular cluster (GC) spatial distribution, which seems inconsistent with the fact that their dynamical friction orbital decay timescales appear to be much shorter than their ages \citep{Tremaine1976,Hernandez1998}. \citet{Goerdt2006} proposed that the problem could be resolved if the structure of the Fornax dark halo had a sizeable constant density core, a result echoed in subsequent work \citep[see; e.g.,][]{Read2006,Cole2012,Petts2015}.

More recent work, however, has argued for a different explanation that does not require a core. \citet{Meadows2020}, for example, report that GCs have similar dynamical friction timescales  in both cuspy {\it and} cored halos normalized to match observed constraints on $M_{\rm tot}(<r_{1/2})$. More precisely,  in either case GCs are driven by dynamical friction to well-defined terminal radii in about the same time. The difference is where they end up: GCs are driven close to the centre in the case of a cusp but ``stall'' at about one-third of the core radius\footnote{The core radius is defined here as the distance where the projected density of the halo drops to $1/2$ of its central value. Since this convention is not always followed, care is needed when comparing quantitative results for the ``stall radius'' from different authors.}  ($\sim 300$ pc for a $\sim 1$ kpc core).
%as they approach a distance to the dwarf centre corresponding roughly to the dark matter core radius (\citet{Goerdt2006} find that %GCs stall at a radius similar to the size of the constant-density region of the cored subhalo, defined as the radius where $|\mathrm{d} %\log \rho / \mathrm{d} \log r |< 0.1$. Applying the same definition of core radius to the ``large core'' model studied in %\citet{Cole2012}, these authors find that GCs stall at a radius of roughly twice the dark matter core radius). }
Because there is no clearcut evidence for such characteristic common radius for GCs in Fornax, this suggests that, if dynamical friction has indeed affected GC orbits, its effects have been mild and might still be ongoing.

In this interpretation, Fornax GCs were likely formed or accreted into Fornax on orbits with radii somewhat larger than where they are today \citep[][]{Angus2009,Boldrini2019}. Indeed, dynamical friction timescales depend sensitively on orbital radius, and it is relatively easy to accommodate the present-day configuration of GCs if their initial radii were just outside Fornax's half-light radius, $r_{1/2}$ \citep{Meadows2020}. Would such clusters survive stripping by Galactic tides if the latter are indeed responsible for the low dark matter content of Fornax?

A further alternative is that GC orbital decay may have been affected by the ``dynamical stirring'' and halo mass loss that could arise from Galactic tides \citep{Oh2000}. If so, the effect of Galactic tides could, in principle, serve to solve the two problems at once, reconciling the low dark matter content of Fornax, as well as the unexpectedly long decay timescales of its GCs, with the predictions of LCDM.

We explore these questions here using a series of controlled, high-resolution \textit{N}-body simulations of the tidal evolution of Fornax. This paper is structured as follows: in section \ref{sec:methods} we discuss our numerical setup, with section \ref{sec:host} detailing our Milky Way host model, section \ref{sec:orbits} describes our choice of orbital parameters, and sections \ref{sec:Fornax} - \ref{sec:code} detail our halo model and \textit{N}-body realisations. Section \ref{sec:results} presents our results, beginning with the dynamical evolution of dark matter in section \ref{sec:DM} and ending with the effect of tides on the stellar component (section \ref{sec:stars}) and globular clusters (section \ref{sec:GC}). We conclude with a summary of our conclusions in Sec.~\ref{sec:conclusions}.

\section{Numerical Setup}
\label{sec:methods}

This section outlines the numerical setup of the \textit{N}-body simulations used to follow the tidal evolution of our Fornax model in the gravitational potential of the Milky Way.

\subsection{Galaxy model}
\label{sec:host}

The Milky Way is modelled as an analytical, static potential, consisting of an axisymmetric two-component \citet{Miyamoto1975} disk, a \citet{Hernquist1990} bulge and an NFW dark matter halo. The model parameters (summarized in Table~\ref{tab:MW}) are as in \citet{Errani2020}, chosen to approximate the \citet{McMillan2011} model with circular velocity $V_\mathrm{c} = 240$\,km\,s$^{-1}$ at the solar circle $R_0 = 8.29$\,kpc. The thick and thin disks are each parametrized by a disk mass $M$, a scale length $a_{\rm d}$ and a scale height $b_{\rm d}$; similarly, the bulge is described by a total mass $M$ and scale length $a$. The NFW model for the Milky Way dark matter halo can be characterized by a scale radius $r_{\rm s}$ and the mass enclosed within that radius, $M_{\rm s}$. These parameters are listed in Table~\ref{tab:MW}.

\subsection{Orbits}
\label{sec:orbits}

The orbit of Fornax in the Milky Way potential is specified by its present-day Galactocentric position and velocity, as inferred from its sky position, radial velocity, distance, and proper motion. Of these, the sky position and radial velocity $v_r = 55.3 \pm 0.3$\,km\,s$^{-1}$ have negligible uncertainties \citep[see; e.g.,][]{Fritz2018} . The heliocentric distance and proper motions, on the other hand, are known to $\sim 10\%$ accuracy: $d = 147 \pm 12.0$\,kpc \citep{Pietrzynski2009}, and $\mu_{\alpha^*} = 0.374 \pm 0.035$ mas/yr, $\mu_\delta = -0.401 \pm 0.035$ mas/yr \citep{Fritz2018}\footnote{Where with $\mu_{\alpha^*}$ we designate the proper motion in $\alpha$ including the $\cos(\delta)$ factor.}.

\begin{figure}
	\includegraphics[width=0.9\columnwidth]{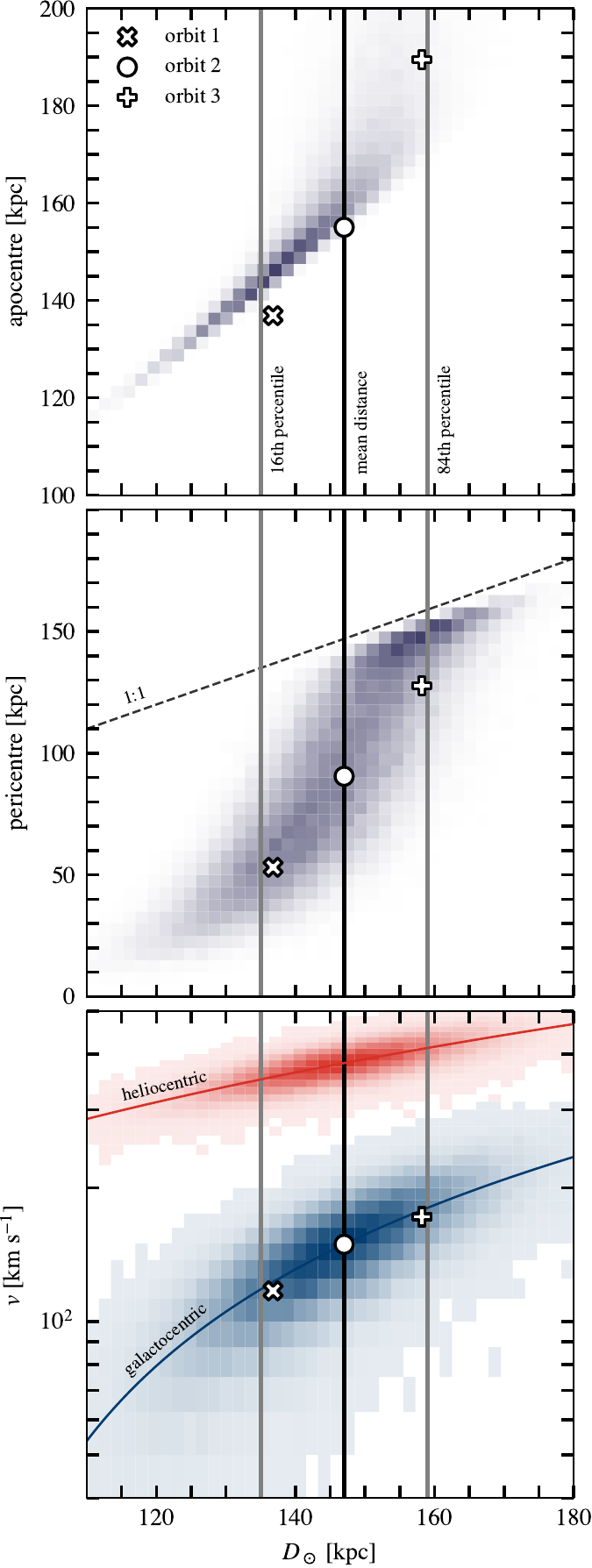}
        \caption{The apocentric (top panel) and pericentric (central panel) distance of Fornax as a function of heliocentric distance, varying the proper motions and radial velocity within the observed uncertainties. The properties corresponding to the three orbits of Table \ref{tab:Fornax_orbit} are highlighted in each panel using white symbols (x$,\circ,+$). Vertical lines correspond to the median measured distance (black) as well as the 16th and 84th percentile (grey).  The bottom panel shows the heliocentric (red) and galactocentric (blue) tangential velocities. Note that the distance uncertainty propagates to a large uncertainty in the Galactocentric velocity. }
    \label{fig:tangential_velocity}
  \end{figure}

These uncertainties result in a fairly broad distribution of possible pericentric and apocentric distances. This is shown in Fig.~\ref{fig:tangential_velocity}, where we plot in the top two panels, as a function of the assumed distance, the resulting pericentric and apocentric distances in the potential described in Sec.~\ref{sec:host}. The pericentric distance, the parameter most critical in respect to the effect of Galactic tides, ranges from $\sim 50$\,kpc at the 16th percentile to $\sim 130$\,kpc at the 84th percentile of the distribution obtained by varying all parameters within their uncertainty range. In other words, a $\sim 10\%$ uncertainty in distance translates into more that a factor of two uncertainty in pericentric distance. 

This striking sensitivity of the pericentric distance on assumed distance is due to Fornax's direction of motion, which lies approximately anti-parallel to the solar motion around the Galaxy. As a result, Fornax's velocity in the heliocentric frame is  mainly tangential and quite large, $\sim 350$\,km\,s$^{-1}$, far exceeding  Fornax's total speed in the Galactocentric reference frame (see Table \ref{tab:Fornax_orbit}).
\begin{table*}
	\centering
	\caption{Current observational constraints, as well as parameters of the three orbits explored using $N$-body simulations in this study. Orbit 2 is the orbit corresponding to the median observed quantities. 
	The current positions and velocities of Fornax corresponding to orbits 1, 2 and 3, expressed as 6D Cartesian coordinates, lie each within at most one standard deviation of the ones derived by \citet{Helmi2018}. Pericentres and apocentres are computed for the Milky Way potential model discussed in section \ref{sec:host}, while heliocentric and galactocentric coordinates are computed with solar parameters of \citet{Schoenrich10}.}
	\label{tab:Fornax_orbit}
	\begin{tabular}{lcccccc}
    \hline
		 \bf{observation}                & $\alpha$        & $\delta$           & distance          &     $\mu_{\alpha^*}$  & $\mu_\delta$        & $v_r$\\
                    		 &  &  & (kpc) & (mas\,yr$^{-1}$) & (mas\,yr$^{-1}$) & (km\,s$^{-1}$)\\ \hline
		 & $2^h 39^m 59.3^s~^{(1)}$ & $-34^\circ 26' 57''~^{(1)}$ & $147 \pm 12.0~^{(2)}$    &     $0.374 \pm 0.035~^{(3)}$ &  $-0.401 \pm 0.035~^{(3)}$ & $55.3 \pm 0.3~^{(3)}$\\ \hline
		 & $X$        & $Y$           & $Z$          &     $V_X$  & $V_Y$        & $V_Z$\\
                    		 & (kpc) & (kpc) & (kpc) & (km\,s$^{-1}$) & (km\,s$^{-1}$)  & (km\,s$^{-1}$)\\ \hline
		 heliocentric & $-32.9$ & $-50.9$ & $-134$    &     $29.3$ &  $-377$ & $75.5$\\ \hline
		 galactocentric & $-41.2$ & $-50.9$ & $-134$    &     $40.4$ &  $-125$ & $82.8$\\ \hline
	\multicolumn{7}{c}{	$^{(1)}$ \citet{McConnachie2012} , $^{(2)}$ \citet{Pietrzynski2009}, $^{(3)}$ \citet{Fritz2018} } \\[0.3cm]
		\hline
		\bf{model parameters} & pericentre & apocentre & distance & $\mu_{\alpha^*}$ & $\mu_\delta$ & $v_r$\\
		 & (kpc) & (kpc) & (kpc) & (mas\,yr$^{-1}$) & (mas\,yr$^{-1}$) & (km\,s$^{-1}$)\\
		\hline
    orbit 1 & 53.1 & 137 & 137 & 0.384 & -0.382 & 55.1\\
		{orbit 2} & 90.5 & 155 & $147$ & $0.374$ & $-0.401$ & $55.3$\\
		orbit 3 & 128 & 189 & 158 & 0.390 & -0.376 & 55.4\\
		\hline

	\end{tabular}
      \end{table*}

Since proper motions are measured in the heliocentric frame, an $\sim 8\%$ uncertainty in distance implies a large tangential velocity uncertainty, roughly $21\%$ in terms of the Galactocentric tangential velocity. This is shown in the bottom panel of Fig.~\ref{fig:tangential_velocity}, where we show the effect of distance on the inferred heliocentric and Galactocentric tangential velocities of Fornax. Because the Galactocentric tangential velocity is not tightly constrained, the orbital pericentre is likewise poorly determined.

\begin{figure*}
	\includegraphics[width=\textwidth]{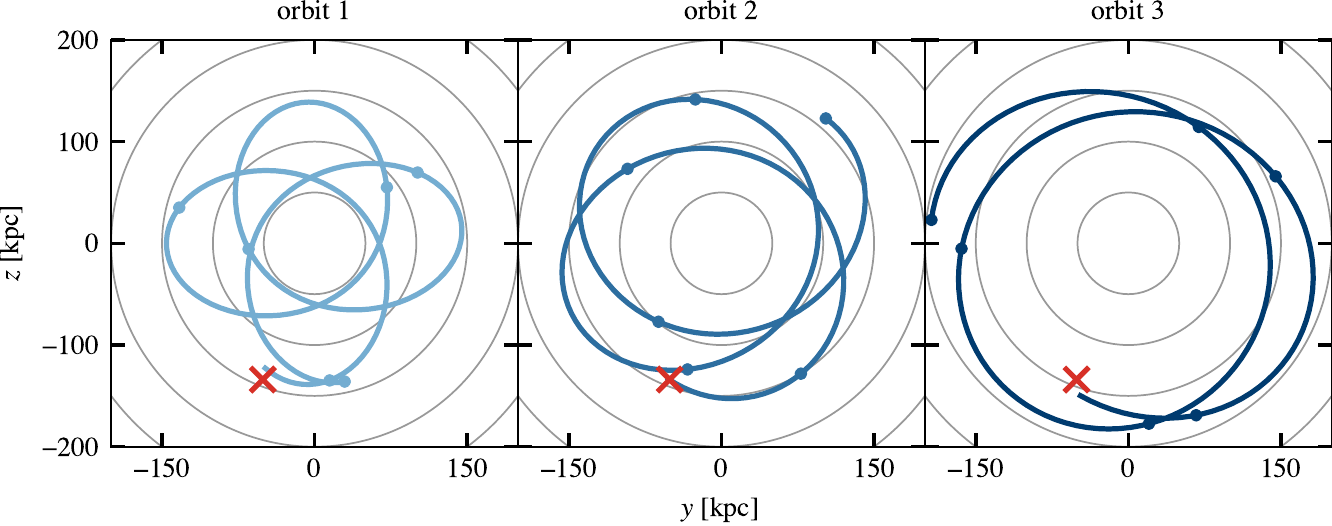}
    \caption{Projected trace on the $Y,Z$ plane of the three orbits considered in this study (see Table~\ref{tab:Fornax_orbit}). The current position of Fornax is indicated by a red cross. Intervals of 2\,Gyr along the orbit are shown using filled circles.}
    \label{fig:yz}
\end{figure*}

Because of this uncertainty, we have explored three different orbits, with orbital properties as given by the open symbols in Fig.~\ref{fig:tangential_velocity}. These orbits have notably different pericentric distances, $r_\mathrm{peri} = 53.1$, $90.5$, and $128$\,kpc, and the corresponding orbits are hereafter referred to as orbits 1, 2 and 3, respectively, with main parameters listed in Table~\ref{tab:Fornax_orbit}. Initial conditions for the $N$-body runs are obtained by integrating those orbits backwards in time for 10\,Gyr. The shape of the resulting orbits are shown in Fig.~\ref{fig:yz}. Note that the three orbits are nearly polar, and close to the $Y$-$Z$ plane\footnote{We use a co-ordinate system where the Sun is located at $(X,Y,Z)_\odot=(-8.3 \mathrm{kpc},0,0)$, and the velocity of the local standard of rest is in the positive $Y$ direction.}. 

\subsection{Fornax model}
\label{sec:Fornax}

We model the Fornax dSph halo as an equilibrium \textit{N}-body realisation of an NFW density profile,
\begin{equation}
    \rho_\mathrm{NFW}(r)=\frac{M_{200}}{4\pi r_s^3}\frac{(r/r_s)^{-1}(1+r/r_s)^{-2}}{[\ln(1+c)-c/(1+c)]}.
	\label{eq:NFW}
\end{equation}
This profile is fully specified by two parameters; e.g., a virial mass, $M_{200}$, and concentration, $c$, or, alternatively, by a maximum circular velocity, $V_\mathrm{max}$, and the radius at which it is achieved, $r_\mathrm{max}$. The latter is often favoured since, unlike the former, it is defined independently of redshift and may be more directly compared with observations.

\begin{figure}
	\includegraphics[width=\columnwidth]{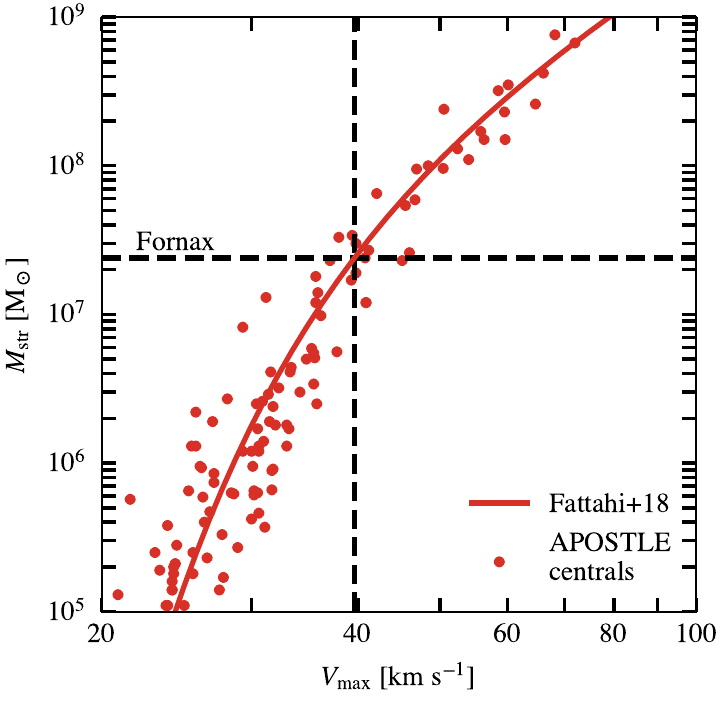}
	\caption{Stellar mass, $M_\mathrm{str}$, versus maximum circular velocity, $V_\mathrm{max}$, of central galaxies 	from the highest resolution level (L1) runs in the APOSTLE simulations as described in \citet{Fattahi2018}. The red solid line is the average relation from that work (Eq.~\ref{eq:Fattahi_fit}). Dashed lines indicate stellar mass and derived $V_{\rm max}$ for Fornax as listed in Table~\ref{tab:Fornax_structure}.}
    \label{fig:APOSTLE}
\end{figure}

Cosmological hydrodynamical simulations have shown that $V_\mathrm{max}$ correlates strongly with galaxy stellar mass, $M_{\rm str}$, as shown, for example, in Fig.~\ref{fig:APOSTLE} for results from the APOSTLE suite of Local Group simulations \citep{Sawala16,Fattahi2016a}.

We use the empirical fit from \citet{Fattahi2018}, 
\begin{equation}
\label{eq:Fattahi_fit}
 M_\mathrm{str} / \mathrm{M}_\odot = m_0 v^\alpha \exp(-v^\gamma)
\end{equation}
where $v = V_\mathrm{max}/50$\,km\,s$^{-1}$, and $(m_0, \alpha, \gamma)=(3.0\times10^8, 3.36, -2.4)$. This fit is shown in Fig.~\ref{fig:APOSTLE}, along with measured stellar masses and maximum circular velocities of isolated galaxies in the APOSTLE simulation.

We adopt a stellar mass for Fornax of $M_{\rm str} = 2.39 \times 10^7\,\mathrm{M_\odot}$, computed from the apparent V-band luminosity as in \citet{Irwin95}, the distance modulus of \citet{Pietrzynski2009} and the stellar mass-to-light ratio of \citet{Woo2008}. Equation \ref{eq:Fattahi_fit} then suggests a maximum circular velocity of $V_\mathrm{max} = 39.6$\,km\,s$^{-1}$. This assumes that tides have not affected Fornax's stellar mass -- an assumption which we shall see justified by our results, as discussed in Sec.~\ref{sec:stars}.

The radius where the  circular velocity peaks, $r_\mathrm{max}$, may be calculated from the \citet{Ludlow2016} parametrisation of the LCDM halo mass-concentration relation at redshift $z=0$. As listed in Table~\ref{tab:Fornax_structure}, the resulting NFW profile has $V_\mathrm{max} = 39.6$\,km s$^{-1}$ and $r_\mathrm{max} = 8.0$\,kpc, or, in terms of virial mass and concentration, $M_{200} = 1.04\times10^{10}$\,M$_\odot$ and $c=12.5$. This initial virial mass is roughly a factor of two lower than the average halo mass inferred for galaxies which host as many GCs as Fornax, but well within the scatter of that relation \citep[see; e.g.,][]{Forbes2018,Shao2020}.

\begin{table}
	\centering
	\caption{Current properties of Fornax ($M_{\rm str}, V_{1/2}, r_{1/2}$) and inferred structural parameters at infall ($V_{\rm max}, r_{\rm max}$). The stellar mass is derived from the distance modulus of \citet{Pietrzynski2009}, the V-band magnitude of \citet{Irwin95}, and the stellar mass-to-light ratio of \citet{Woo2008}. The maximum circular velocity at infall $V_\mathrm{max}$ is estimated from the stellar mass through Eq.~\ref{eq:Fattahi_fit}, and the corresponding $r_\mathrm{max}$ is chosen to match the Ludlow $z=0$ mass-concentration relation \citep{Ludlow2016}. Half-light radius $r_{1/2}$ and circular velocity $V_{1/2}$ at the half-light radius are as in \citet{Fattahi2018}.}
	\label{tab:Fornax_structure}
	\begin{tabular}{ccccc}
		\hline
		$M_{\rm str}$ & $V_\mathrm{max, infall}$ & $r_\mathrm{max, infall}$ & $V_{1/2}$ & $r_{1/2}$\\
		($10^7$\,M$_\odot$) & (km\,s$^{-1}$) & (kpc) & (km\,s$^{-1}$) & (kpc)\\
		\hline
		$2.39$ & $39.6$ & 7.99 & $20.2\pm 2.8$ & $0.949^{+1.06}_{-1.00}$\\
		\hline
	\end{tabular}
\end{table}

The \textit{N}-body realisation of the halo model is constructed with $10^6$ particles ($10^7$ in some cases, as specified in section \ref{sec:GC}) using the \texttt{Zeno}\footnote{\url{http://www.ifa.hawaii.edu/faculty/barnes/zeno/}} software package developed by Joshua Barnes at the University of Hawaii. This software provides routines for Monte Carlo sampling a given distribution function whereby a system in virial equilibrium may be generated. In order for the system to fully relax prior to introducing an external potential, the halo is run in isolation for 6\,Gyr using the publicly available \texttt{GADGET-2} simulation code \citep{Springel2005}, as detailed below.

\subsection{Simulation code}
\label{sec:code}
We use the \textit{N}-body code \texttt{GADGET-2} \citep{Springel2005} to evolve of our \textit{N}-body models. This code implements a hierarchical tree algorithm to compute gravitational interactions. Forces between particles are smoothed with a Plummer-equivalent softening length of $\epsilon_\mathrm{P} = 40$\,pc for $N=10^6$ and $\epsilon_\mathrm{P} = 13$\,pc for $N=10^7$ particles. For each of the orbits described in section \ref{sec:orbits}, the $N$-body model is evolved on the respective orbit in the potential of Section~\ref{sec:host} for $\sim10$\,Gyr. We have also evolved the same N-body models in isolation to identify the smallest radius for which our mass profiles are numerically converged. Defining this as the innermost radius where circular velocities deviate by less than $\sim1\%$ from the target NFW profile, we consider our results converged outside $r_{\rm conv}=450$ pc and $r_{\rm conv}=140$ pc for our $10^6$ or $10^7$-particle runs, respectively.

\section{Results}
\label{sec:results}

We present below the results of the evolution of our Fornax \textit{N}-body model in the Galactic potential for each of the three orbits discussed above. We discuss first the stripping of dark matter, followed by a discussion of the stripping of a hypothetical embedded stellar component that matches Fornax's light distribution. We end with a discussion of how Galactic tides may affect the orbital decay of Fornax GCs.

\subsection{Dark matter stripping}
\label{sec:DM}

\begin{figure}
	\includegraphics[width=\columnwidth]{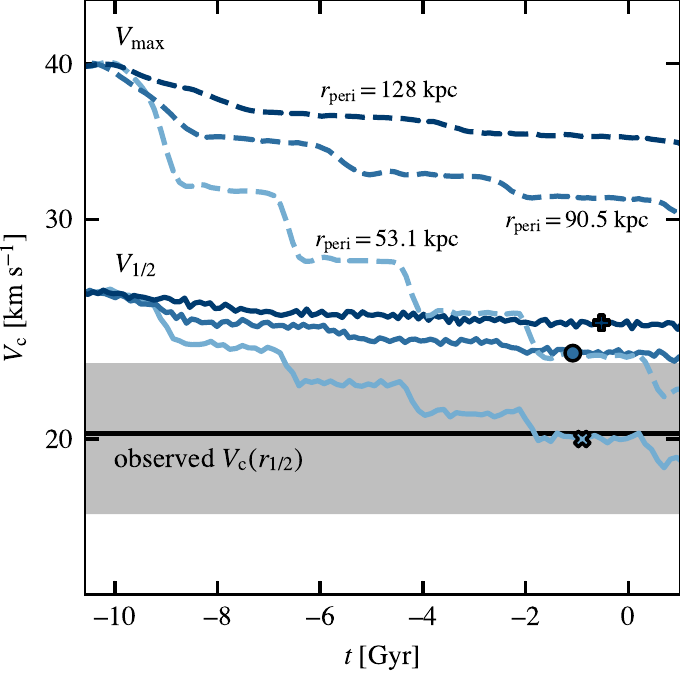}
    \caption{Evolution of the circular velocity of the Fornax halo for the three different orbits, shown in different shades of blue. Lighter shades correspond to smaller pericentres, as labelled. Solid lines show the circular velocity at the half-light radius, $r_{1/2}$, and dashed lines at the radius of maximum circular velocity, $r_\mathrm{max}$. Symbols indicate the snapshot closest to $t=0$ which matches the current galactocentric distance of Fornax. The black solid line and grey shaded region correspond to the observational constraint on $V_{1/2}$ as well as the $\pm 1 \sigma$ uncertainty interval as in \citet{Fattahi2018}.}
    \label{fig:vh}
  \end{figure}

Dashed lines in Fig.~\ref{fig:vh} show the evolution of the maximum circular velocity of our Fornax haloes for the three simulated orbits. Consistent with the findings of \citet{Hayashi03} and \cite{Penarrubia08}, as tides strip the system, $V_\mathrm{max}$ decreases continuously, with abrupt changes corresponding to pericentric passages.
  
As expected, the smaller the pericentre the stronger the tidal forces and thus the more significant the mass loss. For orbit 1 ($r_\mathrm{peri} = 53$\,kpc), the circular velocity at the half-light radius, $V_{1/2} \equiv V_\mathrm{c}(r_{1/2}=1\,$kpc$)$, drops enough after just 4\,Gyr to fall within the 1-$\sigma$ interval around the observational estimate, $20.2 \pm 2.8$\,km\,s$^{-1}$, shown as a grey band in Fig.~\ref{fig:vh} \citep{Fattahi2016b}.  For orbit 1, even after $\sim 10$\,Gyr, $V_{1/2}$ is still consistent with the observed value. The remnant halo in this case retains only $6\%$ of its initial dark mass (yet still fairly well resolved, with more than $2\times10^4$ particles within $r_{\rm max}$), and its $V_{\rm max}$ has dropped from $\sim 40$\,km\,s$^{-1}$ to roughly $23$\,km\,s$^{-1}$. (The snapshots that match best the current Galactocentric distance of Fornax after $\sim 10$ Gyr of evolution are indicated by symbols in Fig.~\ref{fig:vh}.)

On the other hand, models on orbits 2 and 3 ($r_\mathrm{peri} = 91$ and $134$\,kpc) do not lose enough mass for $V_{1/2}$ to reach the 1-$\sigma$ error band of the observationally inferred value, even after $\sim 10$\,Gyr of evolution.

\begin{figure}
	\includegraphics[width=\columnwidth]{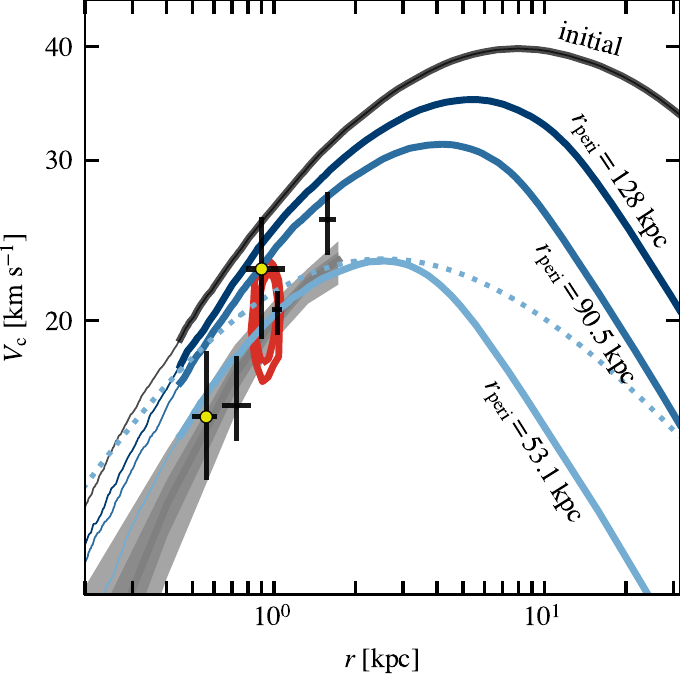}
    \caption{Initial and final circular velocity profiles of the Fornax halo for the three simulated orbits. The red contours are constraints on the enclosed mass within $\sim 1$\,kpc, derived from the stellar velocity dispersion and density profiles (see \citet{Fattahi2016b}, for details). The grey band corresponds to the kinematic analysis of \citet{Read2019}. Error bars with yellow centres show the estimates of \citet{Walker2011}, while the remaining three errorbars show the estimates of \citet{Amorisco2013}. The dotted line shows an NFW profile with $V_\mathrm{max}$ and $r_\mathrm{max}$ as in the final circular velocity profile of orbit 1. The transition from thick to thin lines indicate $r_{\rm conv}$, the convergence radius defined in Sec.~\ref{sec:code}.}
    \label{fig:vc}
\end{figure}

The final circular velocity profiles are shown for the three orbits in Fig.~\ref{fig:vc}. As shown in this figure, the final $V_c$ profile of the tidally-stripped halo on orbit 1 matches well not only the \citet{Fattahi2016b} constraints on the mass enclosed within $\sim 1$\,kpc (red contours), but also -- within 2 sigma -- the mass profile of \citet{Read2019} derived from a kinematic analysis of Fornax’s stellar component (shaded band). Tighter observational constraints at small radii should help to further assess the viability of the tidally-driven scenario proposed here.

Note, in particular, that the mass profile of the stripped halo differs from the NFW shape: the dotted curve in Fig.~\ref{fig:vc} shows an NFW profile with the same $r_{\rm max}$ and $V_{\rm max}$ as the final Fornax model profile on orbit 1 \citep[see; e.g.,][for details]{Errani2021}. The orbit 1 final profile is also in reasonable agreement with mass estimates from \citet{Walker2011} and \citet{Amorisco2013}, shown by the symbols with error bars in Fig.~\ref{fig:vc}.

We conclude that tidal effects may very well reconcile the observed low dark matter content of Fornax with the relatively massive halo suggested by current LCDM cosmological hydrodynamical simulations. This requires that Fornax is on an eccentric orbit with $r_{\rm peri}$ as small as $\sim 50$ kpc, and that it has orbited the Milky Way potential for at least $4$-$5$\,Gyr. This is consistent with the observed proper motions if  Fornax's current distance is of order $140$\,kpc or less, a value well within the error bar on the current estimate of $147\pm 12$\,kpc. Tighter constraints on Fornax's distance would help to verify this prediction. 

\begin{figure*}
	\includegraphics[width=0.95\textwidth]{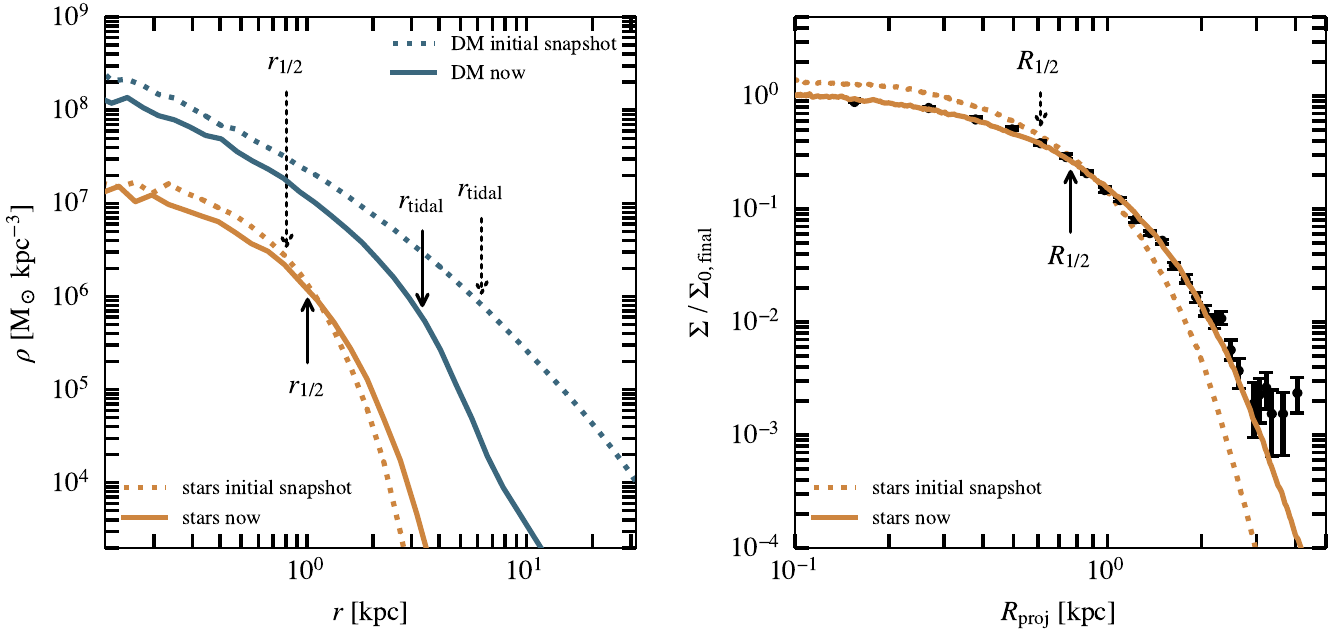}
    \caption{Left: Dark matter (blue) and stellar (orange) density profiles at infall (dotted curves) and after $\sim 10$ Gyr on orbit 1 (solid curves). 3D half-mass and tidal radii are shown using arrows. Right: Surface brightness (orange curves, normalized to the final central value) profiles obtained from the $N$-body simulations, compared against the observational data by \citet{Coleman2005} (black data points). The projected (2D) half-light radii are shown using arrows. The stellar profile, being deeply embedded in the Fornax dark matter halo, is less affected by tides than the dark matter.}
    \label{fig:density}
\end{figure*}

\subsection{Stellar component stripping}
\label{sec:stars}
As discussed in the previous section, Galactic tides are capable of substantially altering the internal distribution of dark matter in the Fornax dSph. We study next to what extent the same tides may affect Fornax's stellar population, limiting this analysis to orbit 1, which is subject to the strongest tides.

To begin, we shall assume that stars are collisionless tracers of the underlying dark matter potential. This is well motivated by the large dark-to-stellar mass ratio of Fornax, as discussed in Sec.~\ref{SecIntro}.
We use the approach introduced by \citet{Bullock2005} to assign stellar probabilities to each dark matter particle, using the publicly available\footnote{\url{https://github.com/rerrani/nbopy}} implementation of \citet{Errani2020}.

The initial stellar density profile is modelled as an Einasto profile \citep{Einasto1965},
\begin{equation}
    \rho_\mathrm{E}(r)=\rho_{\rm E0} \exp\left[-\left( r/a \right)^{1/n}\right],
	\label{eq:Einasto}
\end{equation}
with index $n=0.81$ and scale radius $a=0.44$\,kpc. (The central density $\rho_{\rm E0}$ depends on the stellar mass-to-light ratio assumed, but it should be inconsequential provided the system is dark matter dominated throughout.)

These parameters have been chosen so that after $\sim 10$\,Gyr of evolution, the simulated stellar distribution of the remnant matches the currently observed stellar density profile. The initial and final stellar density profiles are shown in Fig.~\ref{fig:density}. The left panel shows the simulated profiles in 3D, whereas the right panel shows them in projection. These \citet{Sersic1968} profiles fit very well the observed surface density of Fornax (circles with error bars, normalized to the central value) taken from \citet{Coleman2005}.

While tides have a significant effect on the dark matter content of Fornax, the effect of tides on the stellar density distribution is minor. More than $95\%$ of the initial stellar mass remains bound, and its half-mass radius increases only by $\sim 25\%$ over $10$ Gyr of evolution. \citet{Battaglia2015} reached a similar conclusion, wherein more than $99\%$ of stellar particles remain bound, although they assumed a lower initial Fornax halo mass, and orbits with larger pericentric distances. The marginal stellar mass loss corroborates our earlier assumption in section \ref{sec:Fornax}: the current stellar mass of Fornax is essentially unchanged by the Galactic tides, and may thus be used to estimate the halo mass at infall.

\begin{figure*}
	\includegraphics[width=\textwidth]{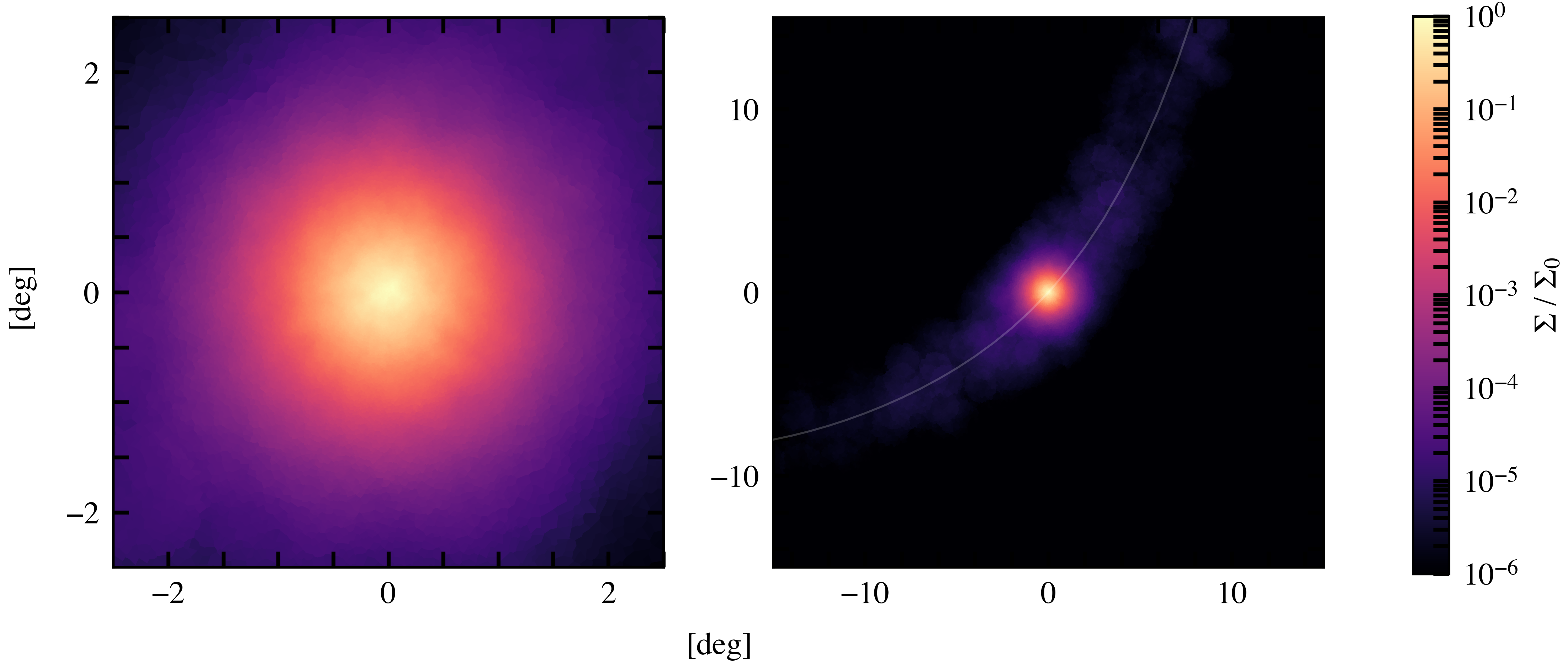}
        \caption{Surface brightness of the evolved Fornax model on orbit 1. At a heliocentric distance of 137\,kpc, one degree in this figure corresponds to 2.4\,kpc. The left panel shows an area of 25 deg$^2$ centred on Fornax. Deviations from spherical symmetry in the surface brightness become only evident in regions where the surface brightness drops below 1/1000 of its central value. The panel on the right shows a larger 900 deg$^2$ areas and suggest the presence of a faint stellar stream. The line passing along the tidal stream in the right-hand panel indicates the orbital path of Fornax.}
    \label{fig:sfb}
\end{figure*}

Fig.~\ref{fig:sfb} shows the distribution of stars in the Fornax model as projected on the sky \footnote{An implementation of the tessellation method used to generate the surface brightness map is publicly available at \url{https://github.com/asya-b/voronoi}.}. The left panel shows an area of $5^\circ \times 5^\circ$ centred on Fornax, and shows no obvious signature of tidal effects, consistent with the recent analysis of  \citet{Wang2019}, who report no tidal signature in the innermost $2^\circ$. Only far from the main body of the galaxy  do signs of tidal distortions become evident. This becomes apparent  when plotting a larger area around the dwarf (right-hand panel of Fig. ~\ref{fig:sfb}),  which shows the presence of a very faint stellar tidal stream, with a surface brightness more  than $1000\times$ fainter than the central surface brightness of Fornax. The total stellar mass outside a sphere of radius 2\,kpc is only $2.6\times 10^6$\,M$_\odot$, or approximately a tenth of Fornax's initial stellar mass.

We conclude that, despite the large losses of dark matter due to stripping, the stellar component remains relatively undisturbed, and should show no obvious signatures of tidal disturbance. Signs of past tidal interaction may be present in the form of very low surface brightness tidal tails around the Fornax dwarf aligned roughly in the direction of the orbital motion, as shown by the curve in the right-hand panel of Fig.~\ref{fig:sfb}. The discovery of such stars would provide strong support for the idea that tidal effects have played an important role in the dynamical evolution of Fornax.

\subsection{Globular cluster orbital decay}
\label{sec:GC}

We now focus our attention on the potential effects of tides on the orbital decay of the five globular clusters associated with the Fornax dSph. \citet{Meadows2020} have recently used $N$-body simulations to model the orbital decay of globular clusters in Fornax, modelled as a cuspy NFW halo. These authors argue that the globular clusters commonly referred to as GC3 and GC4 (see Table~\ref{tab:gcs}) are expected to sink to the centre of Fornax within the next $\sim 2$\,Gyr due to dynamical friction, while the remaining three clusters (GC1, GC2, and GC5) are too far from the centre of Fornax or have too little mass to decay, even in a time span as long as 15\,Gyr. In agreement with earlier work \citep[][]{Angus2009,Cole2012}, these authors also conclude that even if GC3 and GC4 have experienced dynamical friction in a cuspy NFW halo for $\sim 10$\,Gyr, they could have avoided sinking to the centre if their initial orbital radii was $1$-$2$\,kpc.

The models of \citet{Meadows2020}, however, neglect the effect of Galactic tides and assume a relatively low halo mass for Fornax, namely an NFW halo with $V_{\rm max}=25$\,km\,s$^{-1}$, rather than the $40$\,km\,s$^{-1}$ suggested by LCDM simulations. It is therefore important to explore whether their conclusions are robust to changes in the assumed halo mass, as well as to the inclusion of Galactic tides.

To this end we add globular clusters to our Fornax models with $10^7$ particles, including them as single softened point masses, with masses chosen to match the present-day mass of Fornax GCs, between $3.7\times10^4$ and $36\times10^4$\,M$_\odot$ \citep{Mackey2003}. The Plummer-equivalent softening length chosen for all GC particles is $\epsilon_\mathrm{P} = 13$\,pc. Following \citet{Meadows2020}, these clusters are placed on circular orbits at radii between $1$ and $2$\,kpc and evolved first in isolation for 10\,Gyr.

\begin{figure*}
	\includegraphics[width=0.95\textwidth]{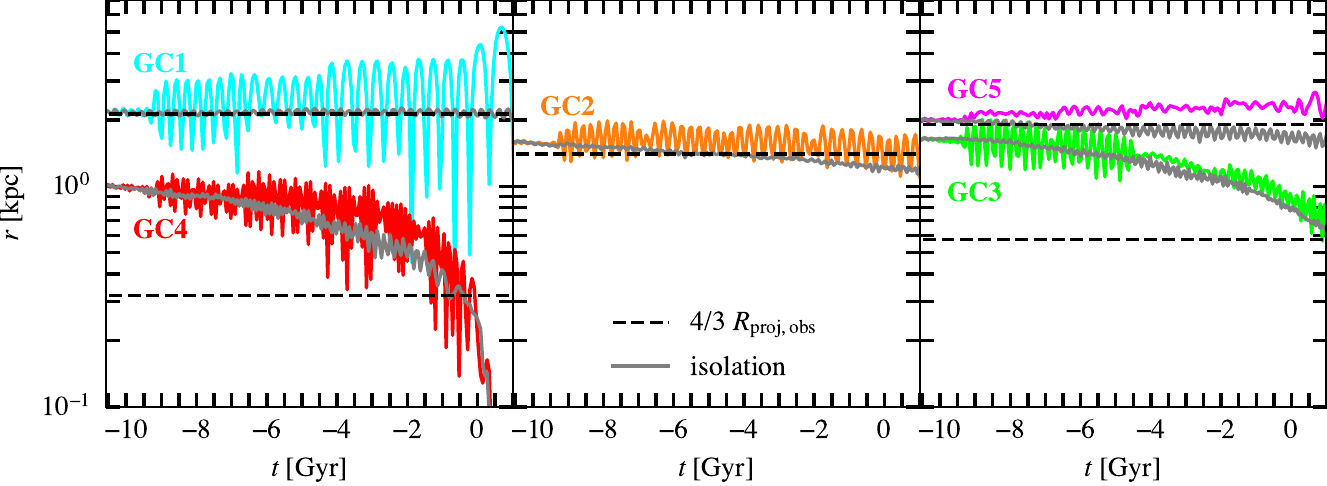}
    \caption{The evolution of the distance to the centre of Fornax for the five GCs. Coloured lines show the evolution in the  presence of tides (orbit 1); grey curves correspond to runs in isolation. Initial radii are chosen so that in absence of tides, the GCs have decayed to approximately their currently observed projected radii in about 10\,Gyr. The main effect of tides is to increase the GC orbital eccentricities (they are initially set on circular orbits); their effect on the orbital decay timescales seems minor at best. }
    \label{fig:gc-t}
\end{figure*}

The orbital decay of these objects in isolation is shown by the grey lines in Fig.~\ref{fig:gc-t}. GC1, GC2, and GC5 evolve very little over 10\,Gyr, so choosing initial orbital radii consistent with their present-day projected radii (indicated by the horizontal dashed lines) results in good agreement with their observed position after 10 Gyr. No appreciable orbital decay is expected for any of these three GCs, at least when Fornax is evolved in isolation.

The situation is different for GC4 and GC3, the two clusters closest (in projection) to the centre of Fornax. Their deprojected 3D distance is shown by the bottom horizontal dashed lines in the left and right panels of Fig.~\ref{fig:gc-t}, respectively. Because of their proximity to the centre and their relatively large masses, GC3 and GC4 decay more quickly, and they need to have started their orbital evolution at $\sim 1.6$\,kpc and $\sim 1.0$\,kpc, respectively, in order to match their present-day position after $\sim 10$ Gyr.

How does including Galactic tides change these conclusions? We explore this by evolving each GC again, but placing the Fornax model in orbit around the Milky Way. We focus here on orbit 1, where the effects of tides are strongest. The resulting evolution of each cluster is shown by the coloured lines in Fig.~\ref{fig:gc-t}. Cluster orbits are significantly affected by Galactic tides; however, the main effect is to increase the GCs' orbital eccentricity, while their orbital decay timescales remain much the same as in isolation.  These results thus suggest that Galactic tides have only a minor effect on the dynamical friction evolution of GCs in Fornax over 10\,Gyr, even in the case of a Galactic orbit with rather small pericentric distance.

The conclusions of \citet{Meadows2020} therefore appear to hold. Only GC3 and GC4 are expected to be affected by dynamical friction; their present-day positions are readily explained if their initial orbital radii was between $1$-$2$\,kpc. This is enough to reconcile the present-day radial distribution of GCs around Fornax with the cuspy dark matter halo profile expected in LCDM.

\begin{table}
	\centering
	\caption{Selected properties of Fornax globular clusters. Both projected radii and masses are taken from \citep{Mackey2003}.}
	\label{tab:gcs}
	\begin{tabular}{lcc}
		\hline
		Name & $R$ & $\log M$\\
		 & [kpc] & [M$_\odot$]\\
		\hline
		GC 1 & 1.60 & $4.57\pm 0.13$\\
		GC 2 & 1.05 & $5.26\pm 0.12$\\
		GC 3 & 0.43 & $5.56\pm 0.12$\\
		GC 4 & 0.24 & $5.12\pm 0.24$\\
		GC 5 & 1.43 & $5.25\pm 0.20$\\
		\hline
	\end{tabular}
\end{table}

\section{Summary and Conclusions}
\label{sec:conclusions}

We have used \textit{N}-body simulations to study the evolution of the Fornax dSph in the Galactic potential. Our main goal is to explore whether Galactic tides may help (i) to explain the relatively low dark matter content of Fornax compared with that expected from LCDM cosmological hydrodynamical simulations of dwarf galaxy formation, and (ii) to reconcile the spatial configuration of its globular clusters with their apparently short dynamical friction orbital decay timescales. Our model assumes that Fornax is a dark-matter dominated system embedded in a cuspy NFW halo. Our main conclusions may be summarized as follows.

Using the latest available data on Fornax's sky position, distance, radial velocity and proper motions, together with a Galactic potential model that matches the most recent dynamical constraints, we conclude that the pericentric distance of Fornax's orbit is only loosely constrained. The main reason for this is the particular direction of Fornax's orbital velocity at present, which is anti-parallel to the Sun's Galactocentric velocity. This implies a large heliocentric tangential velocity, which translates into a sensitive dependence of Fornax's inferred Galactocentric velocity on its assumed distance. Indeed, a $\pm 8\%$ error in the distance translates into pericentric distances that may vary between $\sim 50$ and $130$\,kpc.

Assuming a pericentric distance at the small end of that range, our models suggest that a Fornax NFW halo model with $M_{200}=1\times 10^{10}$\,M$_\odot$ (or, equivalently, $V_{\rm max}=40$\,km\,s$^{-1}$, consistent with the results of the APOSTLE suite of LCDM Local Group simulations) would be stripped of most of its dark mass over $\sim 10$\,Gyr of evolution (over 4 full orbits in the assumed Galactic potential). The tidal mass loss would reduce $V_{\rm max}$ to roughly $\sim 23$\,km\,s$^{-1}$, and the total enclosed mass within half-light radius of Fornax ($r_{1/2}\approx 1$\,kpc) by $\sim 42\%$, bringing it into agreement with observational estimates. 

Our models also indicate that most of the mass lost to stripping is \emph{dark}, and that the total stellar component of Fornax is far less affected. Indeed, a stellar tracer modelled as a S\' ersic  model with $n=0.81$ and $R_{1/2}=0.61$\,kpc (which matches fairly well the observed light profile of Fornax) would be hardly affected, losing less than $5\%$ of its mass in the process. Stripped stars would form tidal tails around Fornax, but with an average surface brightness more than 1000 times fainter than Fornax's central value and therefore extremely challenging to detect. Our models show that the stellar component of Fornax inside at least $\sim 3$\,kpc would show no obvious evidence of any tidal disturbance, again consistent with observations.

Similarly, the effect of Galactic tides on the dynamical friction decay times of Fornax's population of globular clusters is minor compared with models run in isolation. Thus a population of globular clusters with initial orbital radii between $1$ and $2$\,kpc is consistent with their present-day spatial distribution, despite assuming a cuspy NFW halo.

All models implemented here assume tidal stripping in Fornax due to the gravitational influence of the Milky Way only. We note that a massive nearby satellite as the LMC could have a non-negligible contribution to the gravitational potential felt by Fornax, modifying its orbital evolution from that presented here \citep[see e.g.,][]{Erkal2019,Patel2020}. Further uncertainties may arise from the still undetermined total mass of the Milky Way; the total mass considered in this work ($M_{200}\sim 1.15\times 10^{12}$\,M$_\odot$, \citet{McMillan2011}) lies slightly above recent estimates provided by studies based on LMC analogues in cosmological simulations of the Local Group \citep{Santos2021} or stellar halo kinematics \citep{Deason2021}. We plan to consider these effects in future work.

We conclude that both the low dark matter content measured for the Fornax dSph and the radial distribution of its GC population are consistent with a scenario where Galactic tides are solely responsible for stripping a cuspy NFW halo with virial mass $10^{10}\, M_\odot$, as expected from abundance-matching arguments. Note that this scenario does not require a ``core'' but it does require a relatively small pericentric distance ($\sim 50$ kpc) for tides to operate effectively. This is possible if Fornax's true Galactocentric distance is slightly lower than, but within the uncertainty of, current estimates. A tighter distance estimate would therefore provide a helpful check to the validity of this scenario. A further check could be provided by tighter constraints on the innermost mass profile. In the tidal scenario proposed here Fornax's halo is still cuspy; probes of the mass profile within a couple of hundred parsecs would be especially helpful in order to settle questions about the presence of a core or a cusp in the Fornax dSph.

\section*{Acknowledgements}

This work used the DiRAC@Durham facility managed by the Institute for Computational Cosmology on behalf of the STFC DiRAC HPC Facility (www.dirac.ac.uk). The equipment was funded by BEIS capital funding via STFC capital grants ST/K00042X/1, ST/P002293/1, ST/R002371/1 and ST/S002502/1, Durham University and STFC operations grant ST/R000832/1. DiRAC is part of the National e-Infrastructure.

RE acknowledges support provided by a CITA National Fellowship and by funding from the European Research Council (ERC) under the European Unions Horizon 2020 research and innovation programme (grant agreement No. 834148).

AF is supported by a UKRI Future Leaders Fellowship [grant number MR/T042362/1] and Leverhulme Trust.

ISS is supported by the Arthur B. McDonald Canadian Astroparticle Physics Research Institute.

%%%%%%%%%%%%%%%%%%%%%%%%%%%%%%%%%%%%%%%%%%%%%%%%%%
\section*{Data Availability}

The data underlying this article will be shared on reasonable request to the corresponding author.

%%%%%%%%%%%%%%%%%%%% REFERENCES %%%%%%%%%%%%%%%%%%

\bibliographystyle{mnras}
\bibliography{bibliography}

%%%%%%%%%%%%%%%%%%%%%%%%%%%%%%%%%%%%%%%%%%%%%%%%%%

\bsp	% typesetting comment
\label{lastpage}
\end{document}